\documentclass[journal=apchd5,manuscript=article,layout=traditional]{achemso}
\usepackage{amsmath}
\usepackage{float}
\usepackage{color}
\usepackage[normalem]{ulem}
\SectionNumbersOff

\title{Plasmonic Superlens Imaging Enhanced by Incoherent Active Convolved Illumination}
\author{Wyatt Adams}
\author{Anindya Ghoshroy}
\author{Durdu {\"O}. G{\"u}ney}
\email{dguney@mtu.edu}
\affiliation{Department of Electrical and Computer Engineering, Michigan Technological University, Houghton, MI 49931-1295, USA}


\begin{document}

\begin{tocentry}
\includegraphics[scale=1]{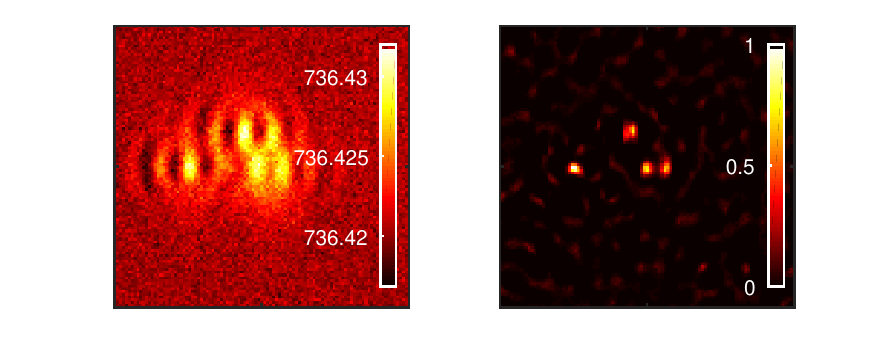}

\vspace{1cm}

Plasmonic superlens imaging with a thin silver film is enhanced by incoherent active convolved illumination method that effectively reduces the losses in the imaging system.
\end{tocentry}

\begin{abstract}
We introduce a loss compensation method to increase the resolution of near-field imaging with a plasmonic superlens that relies on the convolution of a high spatial frequency passband function with the object. Implementation with incoherent light removes the need for phase information. The method is described theoretically and numerical imaging results with artificial noise are presented, which display enhanced resolution of a few tens of nanometers, or around one-fifteenth of the free space wavelength. A physical implementation of the method is designed and simulated to provide a proof-of-principle, and steps toward experimental implementation are discussed.
\end{abstract}

The theory and experimental demonstration of metamaterials has inspired interesting avenues of imaging,\cite{fang_sub-diffraction-limited_2005,taubner_near-field_2006,liu_far-field_2007-1,liu_experimental_2007,smolyaninov_magnifying_2007,fehrenbacher_plasmonic_2015,jacob_optical_2006,liu_far-field_2007,rho_spherical_2010,sun2015experimental,shen2017hyperbolic,zheng2014broadband} lithography,\cite{luo2004surface,gao2015enhancing} and beam generation\cite{liu2017nanofocusing} beyond the diffraction limit, particularly motivated by the prospect of a perfect lens\cite{pendry_negative_2000}. Researchers have quickly realized superlenses\cite{fang_sub-diffraction-limited_2005,taubner_near-field_2006,liu_far-field_2007-1,liu_experimental_2007,smolyaninov_magnifying_2007,fehrenbacher_plasmonic_2015,luo2004surface,gao2015enhancing}, hyperlenses\cite{jacob_optical_2006,liu_far-field_2007,rho_spherical_2010,sun2015experimental}, integrated metalenses,\cite{liu2017nanofocusing,shen2017hyperbolic} and non-resonant elliptical lenses,\cite{zheng2014broadband} which can either amplify or propagate evanescent waves carrying the precious high spatial frequency information of an object. However, these efforts have stopped short of creating a truly perfect lens, since their resolution is limited by losses and they cannot focus light of arbitrary polarization especially at optical wavelengths. Metasurfaces\cite{zhao2011manipulating,aieta2012aberration,kildishev2013planar,pors2013broadband,pfeiffer2013metamaterial,west2014all,khorasaninejad2015achromatic,li2015surface,aieta2015multiwavelength,ma2015planar,yang2016metasurface,genevet2017recent} with unprecedented wavefront manipulation capabilities have also emerged to overcome challenging fabrication and attenuation issues\cite{soukoulis2010optical,soukoulis2011past,guney2009connected,guney2010intra,zhang2015hyperbolic,ndukaife2016plasmonics} relevant to three-dimensional metalenses and other metadevices.

We have previously shown a technique for compensating the loss in a plasmonic metamaterial\cite{aslam_surface_2011} by injecting additional surface plasmon polaritons into the metamaterial via a coupled external beam\cite{sadatgol_plasmon_2015}. This allows the introduction of additional energy into the system to compensate for absorption loss in the metallic structures without detrimentally altering the effective negative refractive index. In contrast to traditional loss compensation methods which often implement gain media\cite{anantha_ramakrishna_removal_2003,noginov_compensation_2008,fang_self-consistent_2009,plum_towards_2009,dong_optical_2010,fang_self-consistent_2010,meinzer_arrays_2010,wuestner_overcoming_2010,ni_loss-compensated_2011,savelev_loss_2013}, the so-called ``plasmon injection" ($\Pi$) scheme does not suffer from the instability introduced by gain or any issues of causality\cite{stockman_criterion_2007}. The desired effective parameters of the metamaterial can then be preserved, and the many practical problems of implementing gain media can be avoided, including the selection of suitable materials and pump sources for specific wavelengths and their limited lifetimes. We subsequently translated the $\Pi$ scheme to superresolution imaging with a homogeneous negative index flat lens (NIFL) with nonzero loss\cite{adams_bringing_2016}, a hyperlens\cite{zhang_enhancing_2016,zhang_analytical_2017}, and a silver superlens\cite{adams_plasmonic_2017}. Our findings showed that linear deconvolution of the image produced by a lossy metamaterial lens is equivalent to physically compensating the loss with injection of an additional structured source, as originally demonstrated with the $\Pi$ \cite{sadatgol_plasmon_2015}. However, passive post-processing can only recover spatial frequency components which are not lost to random noise in the imaging process.

To push the performance of our compensation method further, we developed an active version which relies on the coherent convolution of a high spatial frequency function with an object field focused by a lossy NIFL\cite{ghoshroy_active_2017}. Selective amplification of a small band of spatial frequencies by spatially filtering the object under a strong illumination beam can favorably alter the transfer function so that spatial frequency components which would originally be lost to noise can be successfully transferred to the image plane. In this article, we present a more advanced and versatile method that alternatively employs a simple plasmonic superlens structure illuminated by incoherent UV light, avoiding the complexity and practical difficulties related to phase retrieval or phase detection of coherent fields. Numerical imaging results with this method can resolve point dipole objects separated by a few tens of nanometers, an improvement over passive post-processing. Additionally, we identify that signal-dependent noise provides a limitation to this method and discuss potential strategies for experimental implementation.

\section{Theoretical Description and Noise Characterization}
Incoherent linear shift invariant imaging systems can be conveniently described by the intensity convolution relation
\begin{equation} \label{eq:imaging}
i( \mathbf{r} ) = h( \mathbf{r} ) * o( \mathbf{r} ),
\end{equation}
where $i( \mathbf{r} )$ is an observed intensity image, $h( \mathbf{r} )$ is the incoherent point spread function (PSF) of the system, $o( \mathbf{r} )$ is the object intensity, $\mathbf{r}$ is a position coordinate, and $*$ denotes the convolution operation. Here we will treat the intensities as normalized quantities.
Since convolution in position space is equivalent to multiplication in frequency space, Fourier transformation of eq \ref{eq:imaging} gives the resulting spatial frequency content on the image plane as
\begin{equation}
I( \mathbf{k} ) = H( \mathbf{k} )O( \mathbf{k} ),
\end{equation}
where $\mathbf{k}$ is the spatial frequency and the capital letters denote the respective Fourier transforms. Observing the theoretical and experimental transmission properties of lossy near-field superlenses shows that $H( \mathbf{k} )$ has a low-pass filtering effect on the image\cite{pendry_negative_2000,smith_limitations_2003,podolskiy_near-sighted_2005,moore_improved_2009,moore_experimental_2012}. Consequently, many of the high spatial frequency components of the object are not transferred to the image plane. This is problematic for the imaging of nanometric objects, since the absence of the high-$\mathbf{k}$ information reaching the detection plane can result in an indiscernible blurry image.

To combat the attenuation of this information, we propose a method to recover it by adding additional energy to the system, which we call active convolved illumination (ACI). Consider an "active" object $o_{ACI}( \mathbf{r} )$ obtained by the convolution
\begin{equation}
o_{ACI}( \mathbf{r} ) = o( \mathbf{r} ) * a( \mathbf{r} ) + a_0,
\end{equation}
where $o( \mathbf{r} )$ is the object distribution we wish to obtain, $a( \mathbf{r} )$ is a function that passes high spatial frequencies which we use to inject extra energy to compensate the decaying transmission, and $a_0$ is a ``DC offset'' to ensure that $\forall \mathbf{r}: o_{ACI}( \mathbf{r} ) \geq 0 $. The choice of $a_0$ will be evidently dependent on the term $o( \mathbf{r} ) * a( \mathbf{r} )$, and we select it as a constant for simplicity. However, in a real imaging system, the non-negativity of $o_{ACI}( \mathbf{r} )$ will automatically enforced by the physical propagation, meaning that no knowledge of the object $o( \mathbf{r} )$ would be required. The spatial frequency content is then
\begin{equation} \label{eq:object-convolution}
O_{ACI}( \mathbf{k} ) = O( \mathbf{k} )A( \mathbf{k} ) + a_0 \delta( \mathbf{k} ).
\end{equation}
The relation in eq \ref{eq:object-convolution} is not very informative until we specify the mathematical form of $A( \mathbf{k} )$. Therefore, let us define
\begin{equation}
A( \mathbf{k} ) = 1 + P( \mathbf{k} )
\end{equation}
where $P( \mathbf{k} )$ is of a form convenient in terms of mathematical simplicity and practical considerations, such as a Gaussian function. To perform the ACI compensation, we can then define a series of Gaussian $P( \mathbf{k} )$ and subsequently convolve $A( \mathbf{k} )$ with $O( \mathbf{k} )$. Explicitly, $P( \mathbf{k} )$ can be written as
\begin{equation} \label{eq:P}
P( \mathbf{k} ) = \sum_{j} P_{j} \exp \left[- \frac{( \mathbf{k} - \mathbf{k}_j )^2}{2 \sigma_j^2} \right],
\end{equation}
where $j$ is the Gaussian number, $\sigma_j$ is a parameter proportional to the spatial frequency bandwidth of the $j$th Gaussian, and  $P_{j}$ and $\mathbf{k}_j$ are the amplitudes and center spatial frequencies for the $j$th Gaussian, respectively. Propagation of the ACI object through the superlens system to the image plane gives
\begin{equation}
\begin{split}
&H( \mathbf{k} ) \left[ O( \mathbf{k} )A( \mathbf{k} ) + a_0 \delta( \mathbf{k} ) \right] \\ & = H( \mathbf{k} )O( \mathbf{k} ) + H( \mathbf{k} ) \left[ O( \mathbf{k} )P( \mathbf{k} ) + a_0 \delta( \mathbf{k} ) \right] \\ & = I( \mathbf{k} ) + I_{ACI}( \mathbf{k} ).
\end{split}
\end{equation}
The term $H( \mathbf{k} ) \left[ O( \mathbf{k} )P( \mathbf{k} ) + a_0 \delta( \mathbf{k} ) \right]$ represents the ACI contribution to the image plane spatial frequency content, $I_{ACI}( \mathbf{k} )$. In order to collect deterministic information on the image plane for a selected $\mathbf{k}$ within the bandwidth of $P( \mathbf{k} )$, we then must satisfy the inequality
\begin{equation}
|H( \mathbf{k} )O( \mathbf{k} )A( \mathbf{k} )| > |N( \mathbf{k} )|,
\end{equation}
where $|N( \mathbf{k} )|$ is the noise level. In summary, the reason this method can provide more spatial frequency content than simple passive propagation is that we can control the ``effective'' transfer function so that high-$\mathbf{k}$ components of $O( \mathbf{k} )$ can reach the image plane without being fully attenuated below the noise level, provided that $P_{j}$ is sufficiently large.

To evaluate the prospects of our ACI method in a practical imaging scenario, the effects of noise must be taken into account. Particularly in the case of intensity measurements, since the optical power applied with the active convolution will become large, understanding the signal-dependent nature of the noise is crucial, since the resulting image will possess a substantial mean pixel value. In turn, the signal-dependent noise level will be inherently increased compared with passive imaging. To obtain a noisy image $i_n( \mathbf{r} )$ we assume a parametric noise addition of the form
\begin{equation} \label{eq:noise-model}
i_n( \mathbf{r} ) = i( \mathbf{r} ) + i( \mathbf{r} )^{\gamma} u( \mathbf{r} ) + v( \mathbf{r} ),
\end{equation}
where $i( \mathbf{r} )$ is the noiseless image, $u( \mathbf{r} )$ and $v( \mathbf{r} )$ are independent zero-mean Gaussian random variables, and $\gamma$ is a parameter satisfying $| \gamma | \leq 1$. In this case, we take each term of eq \ref{eq:noise-model} to represent the corresponding photon counts read out by the detector. Additionally, in all following calculations, the photon counts and their corresponding standard deviations are normalized to the maximum count value of the original object distribution that we wish to image. We maintain the same notation for $i( \mathbf{r} )$ as eq \ref{eq:imaging} since we treat both the intensities and photon counts are normalized. Due to the independence of $u( \mathbf{r} )$ and $v( \mathbf{r} )$, we can then write the standard deviation of $i_n( \mathbf{r} )$ as
\begin{equation} \label{eq:std-dev-full}
\sigma_n = \sqrt{ i( \mathbf{r} )^{2 \gamma} \sigma_u^2 + \sigma_v^2},
\end{equation}
where $\sigma_u^2$ and $\sigma_v^2$ are the variances of $u( \mathbf{r} )$ and $v( \mathbf{r} )$, respectively. For our ACI method, $i( \mathbf{r} )$ will become large such that $v( \mathbf{r} )$ is negligible, assuming $\sigma_u$ and $\sigma_v$ have similar orders of magnitude. Therefore, we can simplify eq \ref{eq:std-dev-full} to be
\begin{equation} \label{eq:std-dev-simple}
\sigma_n = \sigma_u i( \mathbf{r} )^{\gamma} .
\end{equation}
Let us then define a signal-to-noise ratio (SNR),
\begin{equation} \label{eq:SNR}
\mathrm{SNR} = \frac{i( \mathbf{r} )}{\sigma_n} = \frac{i( \mathbf{r} )}{\sigma_u i( \mathbf{r} )^{\gamma}} = \frac{i( \mathbf{r} )^{1-\gamma}}{\sigma_u}.
\end{equation}
Optical detectors can reach a SNR of around 60 dB\cite{chen_enhanced_2016,ghoshroy_active_2017}. If we select $\gamma = 1$ in eq \ref{eq:SNR}, $\sigma_u$ becomes $10^{-6}$ for a 60 dB SNR. However, for most realistic detectors $\gamma = 0.5$ due to the Poisson distribution of photon noise \cite{heine_aspects_2006}. In this case, eq \ref{eq:SNR} becomes
\begin{equation} \label{eq:SNR-gamma-half}
\mathrm{SNR} = \frac{\sqrt{i( \mathbf{r} )}}{\sigma_u},
\end{equation}
and the SNR is evidently dependent on the signal.

\section{Imaging Simulation}
\begin{figure}
\includegraphics[scale=1]{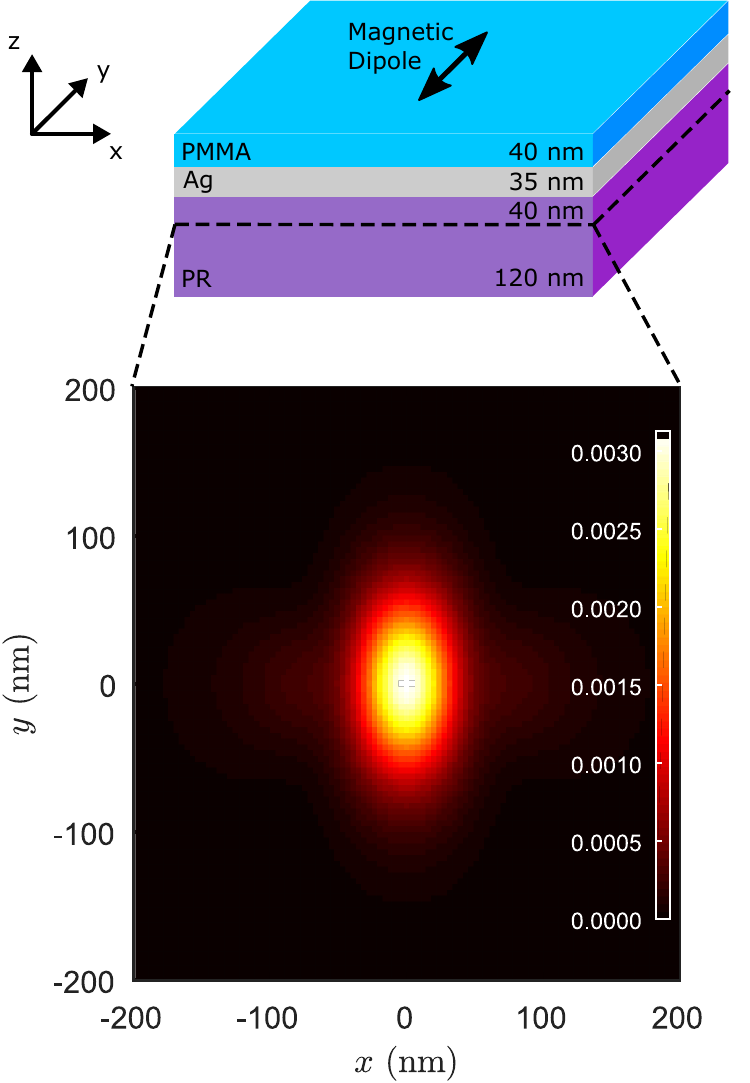}
\caption{Three-dimensional FDTD superlens simulation. A single magnetic dipole embedded in a polymethyl methacrylate (PMMA) dielectric layer is oriented along $y$ and situated 40 nm above the superlens. The image plane (dashed line) in the photoresist (PR) layer is chosen to lie 40 nm below the superlens. The lower plot shows the resulting image plane intensity distribution, which is the PSF of the superlens structure for $y$-polarization of the magnetic field.}
\label{fig:simulation}
\end{figure}

To model a realistic silver superlens structure, we consider a geometry similar to an already experimentally realized superlens which transfers image intensity data onto a photoresist (PR) layer\cite{fang_sub-diffraction-limited_2005}. In order to obtain the PSF for this superlens structure, we simulated the point dipole response with the commercial finite-difference time-domain solver Lumerical FDTD Solutions. The simulation geometry and calculated PSF can be found in Figure \ref{fig:simulation}. In principle, the image plane for a flat silver superlens should lie at the $z$-position where the phase is matched to the object plane. However, there is no such well-defined image plane in Figure \ref{fig:simulation} since the object lies further than a lens thickness from the lens interface on the object side, and the developed PR on the imaging side will have a topographical distribution with varying $z$. Therefore, we somewhat arbitrarily selected an image plane 40 nm from the lens interface in the PR layer for our calculations. This actually highlights that even a defocusing of the image can be overcome with our ACI method. The dipole source is defined as a y-oriented magnetic dipole with center wavelength $\lambda_0 = 365$ nm and bandwidth $\Delta \lambda = 9$ nm to mimic the spectrum of a commercially available UV light-emitting-diode (LED) \cite{noauthor_uv-led/nichia_nodate}. Dispersion in the silver is modeled with a fit to experimental data \cite{palik1998handbook}. The time-average intensity signal in the simulation is finally obtained by an average of the squared modulus of the $y$-component of the magnetic field, $|\mathrm{H}_y(x,y)|^2$, over the bandwidth of the LED source\cite{adams_plasmonic_2017}. In Figure \ref{fig:simulation}, the PSF is asymmetric and narrower along $x$. This is expected, since the Ag superlens has negative permittivity at $\lambda_0 = 365$ nm and can only effectively focus the magnetic field in the direction perpendicular to the dipole orientation. Therefore, illumination with an unpolarized source may slightly worsen the achievable spatial resolution.

\begin{figure*}
\includegraphics[scale=1]{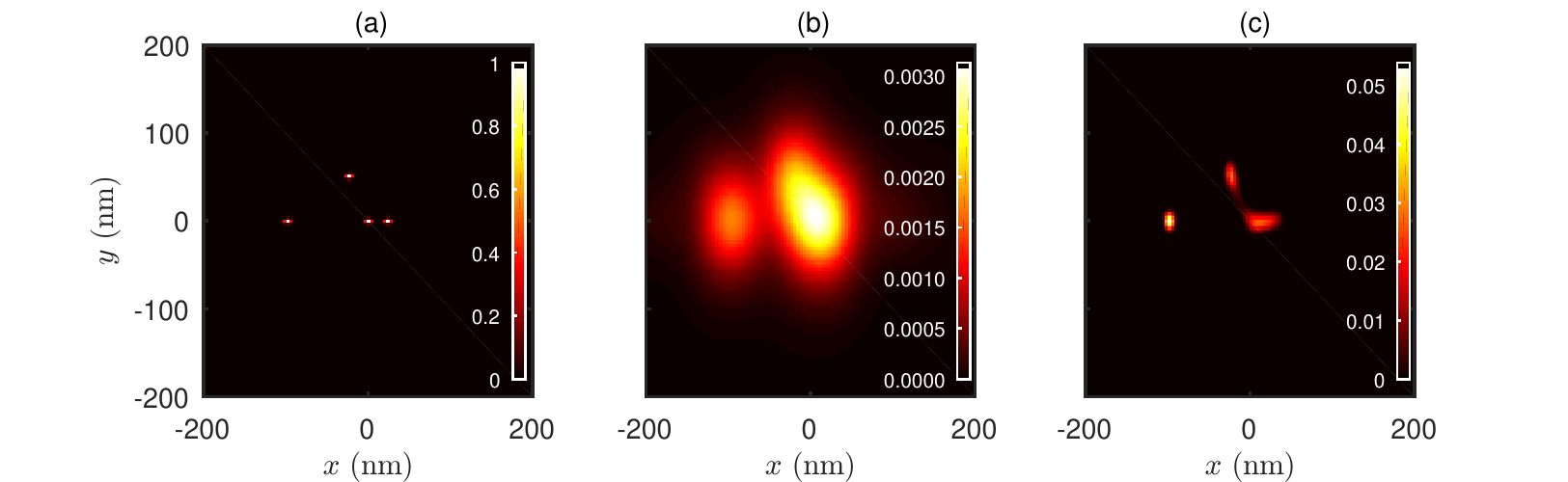}
\caption{Superlens incoherent imaging simulation example of four arbitrarily positioned magnetic point dipole sources. (a) The source distribution at the object plane. (b) Intensity distribution on the image plane. (c) Passive deconvolution of the image in (b) using the Richardson-Lucy algorithm. The two sources near the origin separated by 25 nm are clearly unresolved after deconvolution.}
\label{fig:passive}
\end{figure*}

To obtain the image resulting from a spatially-incoherent distributed object, we can define a distribution of dipole sources on the object plane. Simulating the spatially-incoherent object is then easily performed by adding the contributions of each dipole to the image plane time-average intensity separately. For example, if we consider an object consisting of $n$ dipoles each with intensity $a_j$ and located at position $\mathbf{r}_j$, to obtain the resulting image plane distribution  $i( \mathbf{r} )$ we perform the summation
\begin{equation}
i( \mathbf{r} ) = \sum_{j=1}^{n} \left[h( \mathbf{r} ) * a_j \delta( \mathbf{r} - \mathbf{r}_j ) \right],
\end{equation}
following from the imaging theory in eq \ref{eq:imaging}. An example simulation is shown in Figure \ref{fig:passive} with $n=4$, $a_1=a_2=a_3=a_4$, $\mathbf{r}_1 = (0,0)$ nm, $\mathbf{r}_2 = (25,0)$ nm, $\mathbf{r}_3 = (-100,0)$ nm, and $\mathbf{r}_4 = (-25,50)$ nm. It can be seen that the two sources near the origin, which are separated by a distance of 25 nm, are unresolved even after deconvolution with the iterative Richardson-Lucy algorithm \cite{richardson_bayesian-based_1972,lucy_iterative_1974}. The low-pass filtering due to $H( \mathbf{k} )$ therefore cuts off some of the spatial frequencies that are required for reconstruction of the object. Using our active convolution method, we can recover the lost spatial frequencies that are beyond the cutoff of the passive imaging system.

\section{Results and Discussion}
\begin{figure*}
\includegraphics[scale=1]{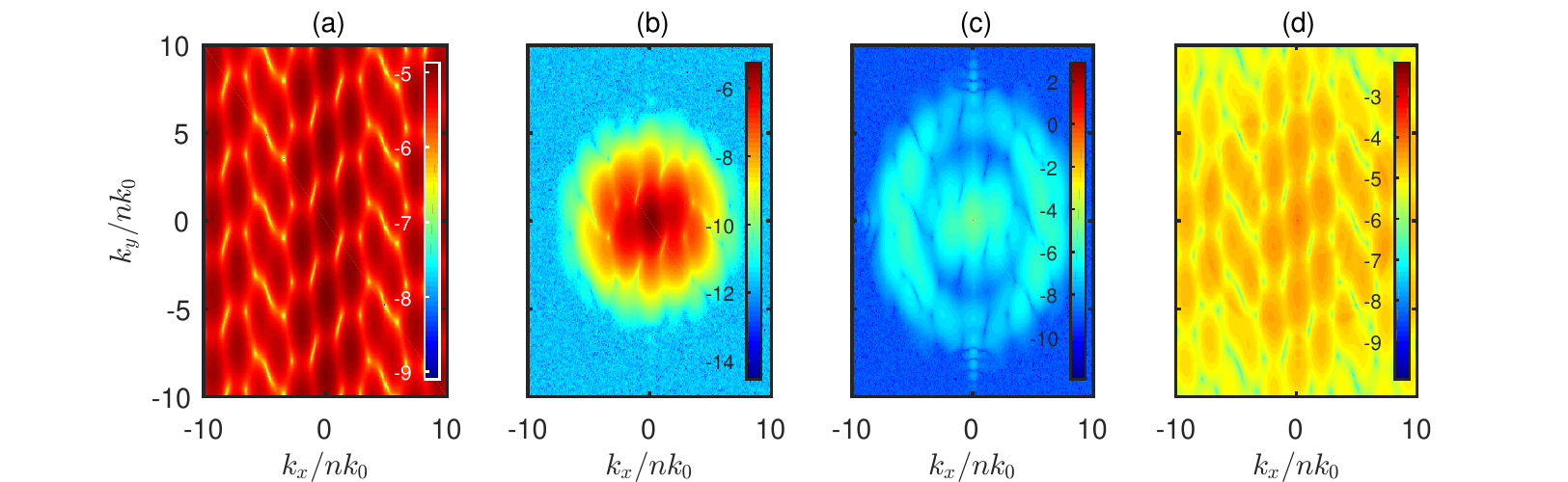}
\caption{Active convolved illumination in the spatial frequency domain. (a) Fast Fourier Transform (FFT) magnitude of the object distribution in Figure \ref{fig:passive} (a). (b) FFT magnitude of the image distribution in Figure \ref{fig:passive} (b). (c) FFT magnitude of the image using the ACI method. (d) FFT magnitude of the image from (c) after deconvolution with the Richardson-Lucy algorithm. All plots are on a logarithmic scale.}
\label{fig:fft}
\end{figure*}

The imaging simulation from Figure \ref{fig:passive} (a) and (b) was used as an example to implement the ACI method. The parameters of $P( \mathbf{k} )$ from eq \ref{eq:P} were chosen to be $P_1 = P_2 = 10^4$, $| \mathbf{k}_1 | = 7 n k_0$, $| \mathbf{k}_2 | = 8 n k_0$, and $\sigma_1 = \sigma_2 = 1.5 n k_0 / 2 \sqrt{2 \log 2}$, where $n = 1.6099$ is the refractive index of the PR imaging medium and $k_0 = 2 \pi / \lambda_0$ is the free space wave number. The resulting ACI object was then propagated through the system using the transfer function $H(k_x,k_y)$ calculated by Fourier transformation of the simulated PSF $h(x,y)$. The resulting spatial frequency content is shown in Figure \ref{fig:fft}. In Figure \ref{fig:fft} (c), it can be seen that a larger band of spatial frequencies are recovered on the image plane compared to (b) due to propagation of the ACI object. After reconstruction with the Richardson-Lucy algorithm and the ``active'' PSF
\begin{equation} \label{eq:active-psf}
h_{ACI}(x,y) = h(x,y) * a(x,y),
\end{equation}
the object spectrum is mostly recovered in Figure \ref{fig:fft} (d). Note that the DC component introduced by the ACI procedure is excluded in eq \ref{eq:active-psf} since it only contributes to the $\mathbf{k}=0$ component and in turn has no effect on the the reconstruction.
\begin{figure*}
\includegraphics[scale=1]{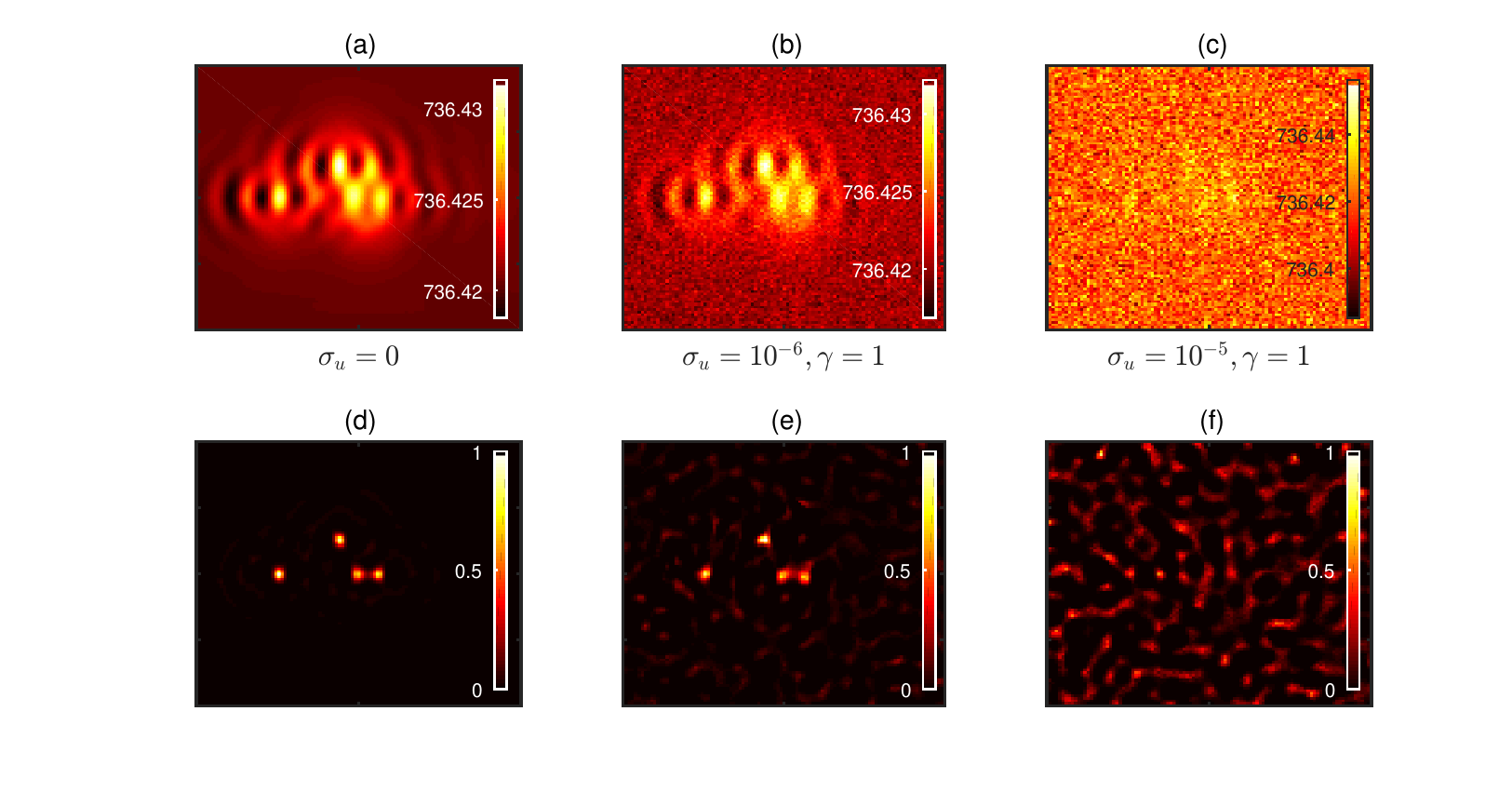}
\caption{Active convolved illumination imaging in the presence of signal-dependent noise. (a)-(c) The calculated images after applying the ACI method for $\gamma = 1$ and different values of $\sigma_u$ from eq \ref{eq:std-dev-simple}. These are the ``measured'' images that would be detected in an experiment. (d)-(f) The final reconstructed images after deconvolution with the Richardson-Lucy algorithm and the active PSF defined in eq \ref{eq:active-psf}. (a) and (d) are the ideal results with $\sigma_u = 0$. Qualitatively, it can be seen that the image in (c) is corrupted by noise, and the reconstructed image in (f) consequently suffers. However, the noise in (b), corresponding to a 60 dB SNR, is small enough to achieve a good reconstructed image in (e).}
\label{fig:gamma-one}
\end{figure*}
\begin{figure*}
\includegraphics[scale=1]{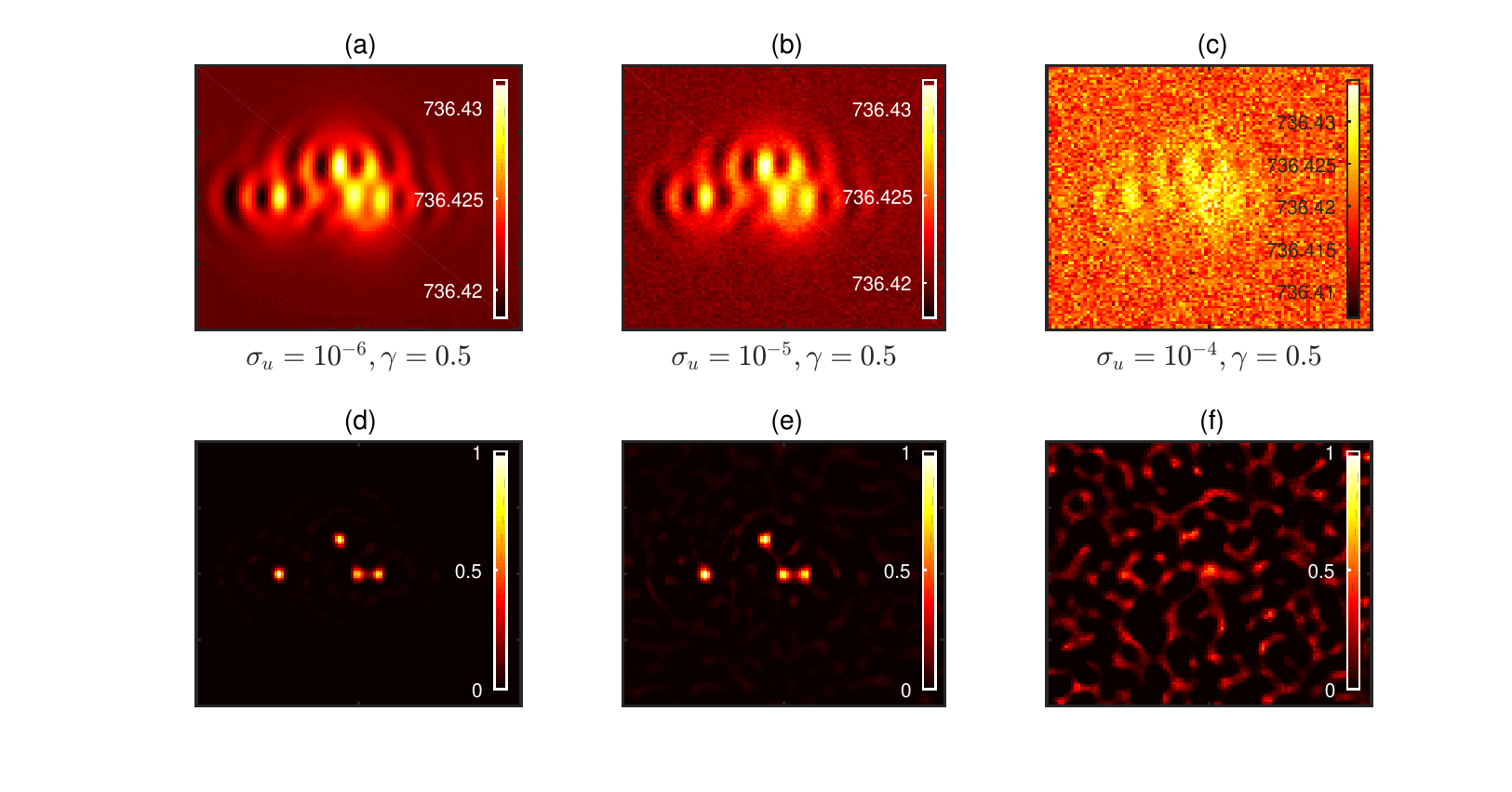}
\caption{Active convolved illumination with $\gamma = 0.5$ and three different values of $\sigma_u$. (a)-(c) The noisy ACI images and (d)-(f) the corresponding reconstructions. (d) and (e) successfully resolve the object, however the image in (c) is too noisy to obtain a good reconstruction in (f).}
\label{fig:gamma-half}
\end{figure*}
\begin{figure}
\includegraphics[scale=1]{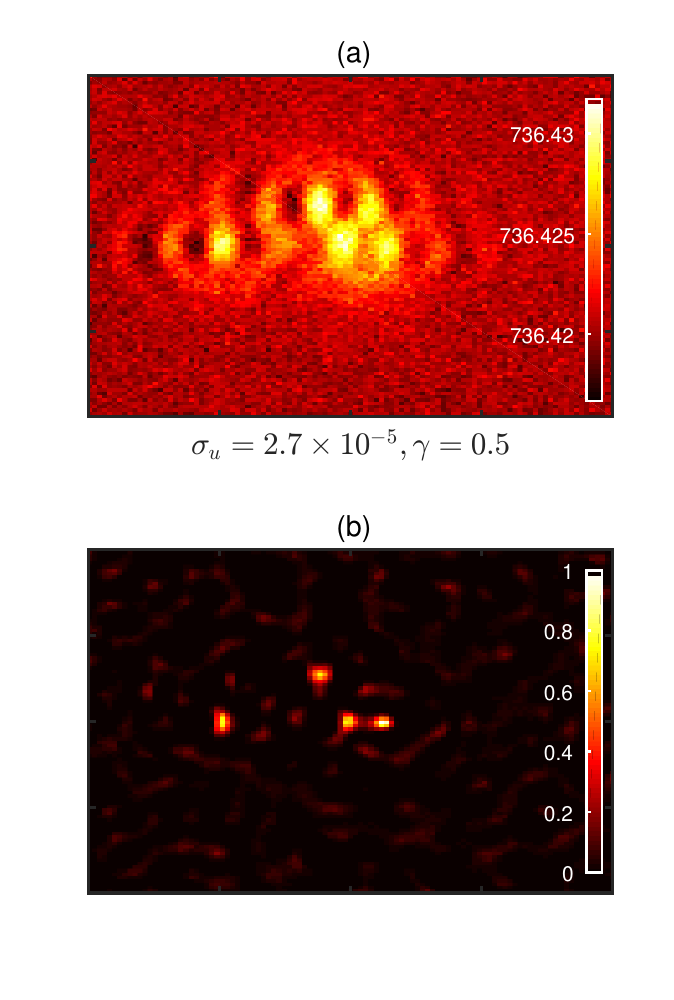}
\caption{Active convolved illumination imaging with $\gamma = 0.5$ and $\mathrm{SNR} \approx 60$ dB. (a) The noisy ACI image and (b) the corresponding reconstruction. In (b) the object is successfully resolved, despite the addition of realistic noise in (a).}
\label{fig:gamma-half-60dB}
\end{figure}
The calculation in Figure \ref{fig:fft} was performed with $\sigma_u = 0$ and $\sigma_v = 10^{-6}$ to better explicate the impact of the ACI on the imaging process. However, it is crucial to evaluate the ACI imaging performance in the presence of signal-dependent noise. To do so, we added simulated noise with $\gamma = 1$, corresponding to a constant SNR, and $\gamma = 0.5$ for a signal-dependent SNR. We have chosen these values of $\gamma$ to represent both the expected Poissonian counting statistics ($\gamma = 0.5$) and a ``worst case" scenario ($\gamma = 1$) to both compare the effects of signal-dependent and signal-independent SNR as well as evaluate the robustness of our method to a variety of noise conditions. Figure \ref{fig:gamma-one} shows the imaging results for $\gamma = 1$ with varied $\sigma_u$. Figure \ref{fig:gamma-one} (a) and (d) respresent the ideal ACI image and corresponding reconstruction with $\sigma_u = 0$. The ACI image in (b) and the corresponding successful reconstruction in (e) consider a 60 dB SNR attainable with modern photodetectors. Unfortunately, decreasing the SNR to 50 dB in (c) leads to an image almost fully corrupted by noise that cannot be reconstructed in (f). In contrast, Figure \ref{fig:gamma-half} considers the case of Poisson-distributed noise with $\gamma = 0.5$ and varied $\sigma_u$. As shown in eq \ref{eq:SNR-gamma-half}, when $\gamma = 0.5$ the SNR becomes dependent on the signal level. Therefore, the $\sigma_u$ value we can define as a ``realistic" noise in Figure \ref{fig:gamma-half} is not explicit. However, if we inspect eq \ref{eq:SNR-gamma-half} and take $i( \mathbf{r} ) \approx \overline{i( \mathbf{r} )}$, where $\overline{i( \mathbf{r} )}$ is the mean pixel value, we can solve for the $\sigma_u$ corresponding approximately to a 60 dB SNR. We can make this approximation since the variations in $i( \mathbf{r} )$ are four orders of magnitude smaller than $\overline{i( \mathbf{r} )}$ for this specific imaging example. The ACI imaging results for these parameters are shown in Figure \ref{fig:gamma-half-60dB}, and it can be clearly seen in (b) that the object can again be successfully resolved.

There are a few aspects of the ACI method that require some qualitative discussion, in particular the limitations of incoherent ACI for increasing the resolution of an imaging system. So-called ``perfect" imaging exhibiting a flat effective transfer function could in principle be approached with this method by iteratively applying multiple $P( \mathbf{k} )$ passing distinct spatial frequency bands so that the full spectrum of the object can be recovered from the noise. However, since the main objective of ACI is to add energy to a narrow band of the object's spatial spectrum in order to overcome the attenuation of those spatial frequencies by the imaging system, it becomes evident that shifting this band to higher $\mathbf{k}$ will require larger intensities. Therefore, we are met with a trade-off between increasing the detectable spatial frequencies and reducing the noise to an acceptable level. We have identified this trade-off as the theoretical limit of the spatial resolution achievable with the ACI method. Using the simulations and parameters described above, we found that the best resolution we could obtain was about 20 nm while considering a 60 dB SNR. However, this could be improved by better optimizing the phase and impedance match between the object and image planes by appropriately tuning the refractive indices of the dielectrics surrounding the superlens along with the locations of each plane. As an example, in our simulation from Figure \ref{fig:simulation}, it is reasonable to expect a better resolution when the dipole is moved slightly closer to the Ag layer, since more of the evanescent components from the source will reach the superlens. We expect these steps would better optimize the results in terms of spatial resolution. Our ACI method could be just as effectively applied to any of these different geometries.

\begin{figure}
\includegraphics[scale=1]{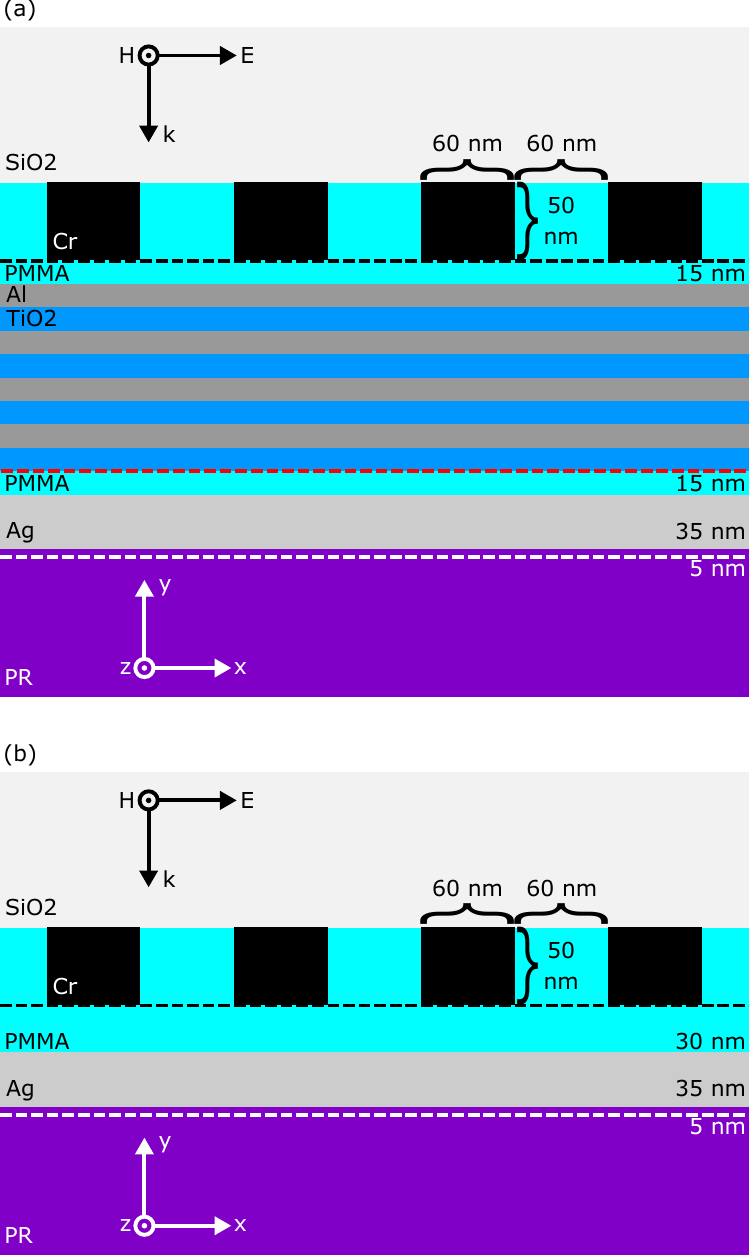}
\caption{(a) Simulation geometry for the physical realization of ACI superlens imaging with a hyperbolic metamaterial spatial filter. The black, red, and white dashed lines indicate the object, active convolution, and image planes, respectively. The image plane is set 5 nm below the Ag superlens in order to make the total propagation distance in the PMMA and PR equal to the lens thickness. (b) The passive simulation geometry used for comparison with the results from (a). The PMMA layer separating the Cr mask and the Ag superlens is set to 30 nm in order to match the total propagation distance in the PMMA in (a).}
\label{fig:grating-HMM-superlens}
\end{figure}

\begin{figure}
\includegraphics[scale=1]{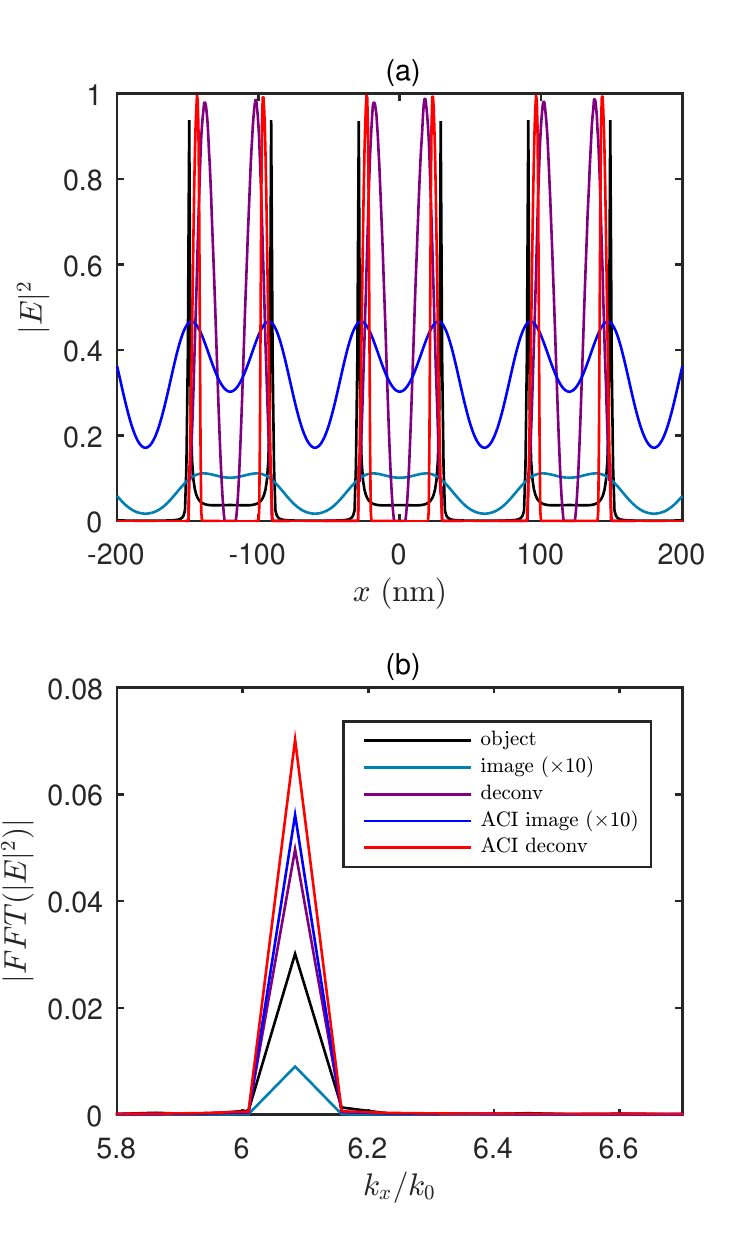}
\caption{(a) Electric field intensities at the object and image planes for the ACI and superlens imaging configurations from Figure \ref{fig:grating-HMM-superlens} considering 60 dB SNR, along with the reconstructions calculated with the Richardson-Lucy deconvolution algorithm. Deconvolution of the ACI image (red solid line) better matches the sharp peaks in the object plane (see black solid lines) as compared to deconvolution of the image formed by Figure \ref{fig:grating-HMM-superlens} (b) (see purple solid line). (b) FFT magnitudes of the data in (a) within the passband of the HMM spatial filter. A clear enhancement of the spatial frequency content near $k_x/k_0=6.1$ can be seen for the ACI image (blue solid line) as compared to the image formed by the silver superlens alone (turquoise solid line). For the ACI image, the ratio of this spectral component to the DC component is increased by more than $18\%$ over the unfiltered superlens image.}
\label{fig:grating-HMM-superlens-imaging-results}
\end{figure}

The most pressing obstacle for implementation of the ACI method into an experimental system is creating the physical convolution in eq \ref{eq:object-convolution}. One way to do this would be to illuminate the object with a high-intensity beam and then spatially filter the near-field intensity distribution. We have shown the design of a near-field spatial filter\cite{ghoshroy_hyperbolic_2017} based on hyperbolic dispersion for a similar function to approximate the behavior of $P( \mathbf{k} )$ in eq \ref{eq:P} and used the filter under ``coherent" convolved illumination for enhanced superlens imaging\cite{ghoshroy_enhanced_2018}. We can use a similar configuration for realization of ACI even when we do not have access to the phases of the fields and the light is not strictly perfectly coherent. Suppose we want to image the intensity pattern formed by a periodic Chromium (Cr) grating object illuminated with TM-polarized light as shown in Figure \ref{fig:grating-HMM-superlens}. Here we consider the same LED light with $\lambda_0=365$ nm and $\Delta\lambda=9$ nm as in the previous simulations. To realize the convolution, we place below the grating a hyperbolic spatial filter we have designed which passes a small band of the spatial frequencies near $6k_0$ which are present in the selected grating. The filter is formed by alternating layers of Aluminum (Al) and Titanium dioxide (TiO\textsubscript{2}) with thicknesses of 16 nm and 15 nm, respectively. Each pair of metal-dielectic layers constitutes a unit cell of the hyperbolic metamaterial (HMM). In this case, we only need to use 4 unit cells to construct the filter since the low spatial frequencies which would otherwise tunnel through the filter are automatically rejected by the grating. This also has the positive side effect of better transmission compared to the corresponding filter we previously designed\cite{ghoshroy_enhanced_2018}. The red dashed line in Figure \ref{fig:grating-HMM-superlens} (a) at the exit interface of the HMM is the plane at which the active convolution exists. To tie the physical system in with our theory, the HMM spatial filter essentially performs the operation in eq. \ref{eq:object-convolution}, and the amplitude of $P( \mathbf{k} )$ can be controlled by simply modulating the intensity of the illumination incident on the grating object. The ACI image is then formed at the image plane (white dashed line) within the PR layer. Using FDTD solutions, we performed simulations of this geometry, and also the passive configuration in Figure \ref{fig:grating-HMM-superlens} (b), in order to provide a physical proof-of-concept for our ACI method and its potential for enhancing the resolution. In this case we increased the illumination intensity by selecting $P_1 = 10^8$ to fully overcome the added 60 dB signal-independent noise, but decreasing this value by about two orders can still give good results. A 10-20 dB signal-dependent SNR was also found to be tolerable using these parameters. The simulation results can be found in Figure \ref{fig:grating-HMM-superlens-imaging-results}. The intensity profile induced on the Cr grating mask shows ``hot spots" that occur every 60 nm at the sharp edges of the grating (see black solid line  in Figure \ref{fig:grating-HMM-superlens-imaging-results} (a)). The spatial frequency corresponding to a 60 nm period is within the passband of the HMM spatial filter, and in Figure \ref{fig:grating-HMM-superlens-imaging-results} (a) this frequency is accentuated in the ACI image (blue solid line) compared to the passive image (turquoise solid line). The magnitudes of this frequency for the data in (a) are shown in (b) for comparison. Finally, the deconvolution of the ACI image with the PSF (calculated by removing the grating and placing a point source on the object plane) gives a better representation of the intensity induced on the grating than the passive deconvolution (i.e., compare the red solid line with the purple). This result can in principle be improved by tuning the spatial filter to higher spatial frequencies, provided that it passes one of the primary grating frequencies.

It is interesting to note that such spatial filters integrated with a superlens cavity and type I hyperbolic metamaterial have also been recently proposed for nanofocusing of Bessel beams\cite{liu2017nanofocusing} and  an implementation of a hyperbolic dark-field lens\cite{shen2017hyperbolic}, respectively.  Additionally, spatial filtering has been shown to reduce the line edge roughness of photolithographic exposures in the presence of surface roughness\cite{liang2018achieving}. Therefore, one natural extension of our work would be the studying of these intriguing high-resolution imaging systems from the ACI method perspective to not only experimentally confirm our theoretical predictions but also improve their performances. The actual transmission properties of the filter may not exactly replicate $P( \mathbf{k} )$ from eq \ref{eq:P}, but the important property is the ability to selectively amplify a band of high spatial frequencies relative to the spatial frequencies in the original passband of the superlens. We showed in Ghoshroy \textit{et al.}\cite{ghoshroy_active_2017} that simply amplifying the entire object spectrum (the ``strong illumination" case) will only result in a deleterious amplification of the signal-dependent noise in the image. This is why the selective amplification of a finite portion of the spatial spectrum is required. A secondary obstacle for experimental implementation is the near-field detection of the subwavelength intensity distributions produced by this method. Near-field images produced by silver superlenses operating in the UV are often read out by exposing a negative tone PR layer on the imaging side of the lens, developing the PR, then characterizing the developed PR topography with an atomic force microscope. The ACI imaging system we have shown in Figure \ref{fig:grating-HMM-superlens} (a) is within the capabilities of modern nanofabrication. However, it is likely that scaling the experiment to infrared, terahertz, or microwave frequencies would be more convenient in terms of both fabrication and detection. There is no theoretical restriction to scaling our ACI method to other frequencies. The images could then be directly read out with a subwavelength near-field probe\cite{inouye_near-field_1994,hillenbrand_pure_2001,hillenbrand_material-specific_2002,gerton_tip-enhanced_2004,taubner_near-field_2006,hoppener_antenna-based_2008,kehr_near-field_2011,rudolph_broadband_2012,fehrenbacher_plasmonic_2015,kehr_local_2016,adams_review_2016} or detector\cite{kim_simulation_2008,elkhatib_subwavelength_2009,gregoire_sub-wavelength_2009,kawano_highly_2011,bergeron_introducing_2012,bergeron_diffraction_2013, inampudi_diffractive_2013,mitrofanov_photoconductive_2015,fridental_image_2016,mitrofanov_near-field_2017} small enough to resolve the important features.

\section{Conclusion}
We developed a loss compensation method to improve the resolution of a near-field silver superlens using incoherent active convolved illumination. A theoretical description of the imaging method for incoherent light is developed and implemented in numerical simulations using a combination of the finite-difference time-domain method and linear shift-invariant imaging theory. The presence of signal-dependent noise is taken into account to represent a realistic imaging scenario. The imaging method presented can achieve a resolution of around $\lambda_0/15$ or better under optimal phase and impedance matching conditions even when corrupted by realistic noise. The theory was then implemented in the design and simulation of a superlens imaging system that uses a hyperbolic metamaterial spatial filter to perform the required convolution operation physically to improve the imaging performance. The results do not only indicate the power of superlenses for enhanced sub-diffraction imaging but also the efficacy of the $\Pi$ loss compensation scheme by decently connecting and attempting to resolve two grand issues of optics, namely loss compensation and imaging beyond diffraction limit. The experimental implementation of the imaging method was also discussed.

\begin{acknowledgement}
This work was supported by Office of Naval Research (award N00014-15-1-2684).
\end{acknowledgement}

%\bibliography{ACS-Photonics-submission}
\providecommand{\latin}[1]{#1}
\makeatletter
\providecommand{\doi}
  {\begingroup\let\do\@makeother\dospecials
  \catcode`\{=1 \catcode`\}=2 \doi@aux}
\providecommand{\doi@aux}[1]{\endgroup\texttt{#1}}
\makeatother
\providecommand*\mcitethebibliography{\thebibliography}
\csname @ifundefined\endcsname{endmcitethebibliography}
  {\let\endmcitethebibliography\endthebibliography}{}

\end{document}